\documentclass[twocolumn]{aastex61}
\usepackage{hyperref}
\usepackage{amsmath}
\usepackage{afterpage}

\usepackage{soul} 
\usepackage[caption=false]{subfig}
\shorttitle{Disentangling Dark Physics with CMB}
\shortauthors{Li et al.}

\begin{document}
\title{Disentangling Dark Physics with Cosmic Microwave Background Experiments}
\correspondingauthor{Zack Li}
\email{zequnl@astro.princeton.edu}
\author[0000-0002-0309-9750]{Zack Li}
\affil{Department of Astrophysical Sciences, Princeton University\\
4 Ivy Lane \\
Princeton, NJ 08544, USA}
\author{Vera Gluscevic}
\affiliation{School of Natural Sciences, Institute for Advanced Study\\
Einstein Drive, Princeton, NJ 08540, USA}
\author{Kimberly K. Boddy}
\affiliation{Department of Physics \& Astronomy, Johns Hopkins University\\ Baltimore, MD 21218, USA}
\author{Mathew S. Madhavacheril }
\affil{Department of Astrophysical Sciences, Princeton University\\
4 Ivy Lane \\
Princeton, NJ 08544, USA}

\begin{abstract}
We forecast constraints on dark matter (DM) scattering with baryons in the early Universe with upcoming and future cosmic microwave background (CMB) experiments, for DM particle masses down to 15 keV.
In terms of the upper limit on the interaction cross section for a velocity-independent spin-independent elastic scattering, compared to current \textit{Planck} results, we find a factor of $\sim$6 improvement with CMB-Stage 3, a factor of $\sim$26 with CMB-Stage 4, and a factor of $\sim$200 with a cosmic-variance limited experiment.
Once the instrumental noise reaches the proximity of 1 $\mu$K-arcmin, the constraints are entirely driven by the lensing measurements.
The constraints benefit from a wide survey, and show gradual improvement for instrumental noise levels from 10 $\mu$K-arcmin to 1 $\mu$K-arcmin and resolution from 5 arcmin to 1 arcmin.
We further study degeneracies between DM interactions and various other signatures of new physics targeted by the CMB experiments.
In the primary temperature and polarization only, we find moderate degeneracy between the effects of DM scattering, signals from massive neutrinos, and from the effective number of relativistic degrees of freedom.
The degeneracy is almost entirely broken once the lensing convergence spectrum is included into the analyses.
We discuss the implications of our findings in context of planned and upcoming CMB measurements and other cosmological probes of dark-sector and neutrino physics.
\end{abstract}

\section{Introduction}

The particle nature of dark matter (DM) is unknown, and many experiments and observations are designed to search for signs of its non-gravitational interactions.
At low energies, most notably, direct DM searches which rely on nuclear-recoil measurements in low-background underground experiments place strong bounds on DM interactions with the Standard Model, for DM particle masses above a GeV \citep{cushman2013, aprile2017}. 
Such experiments will push the bounds to even lower values in the near future \citep{lux2017, cresst2016}.
However, the range of DM masses and cross sections that direct detection can currently probe is limited, and new strategies are being devised to sidestep these limitations \citep{essig2012,essig2017,supercdms2017,zaharijas2005, cresst2016, mahdawi2017, emken2017}.

Cosmological data such as the measurements of the cosmic microwave background (CMB) and large scale structure (LSS) provide insight into DM physics, and are sensitive to high cross sections and the full range of masses down to sub-GeV particles---all of which are inaccessible to most other low-energy experiments.
The main effect of DM-baryon scattering on cosmological observables is suppression of structure at small scales. 
The strongest limits on DM-baryon scattering in the early universe currently come from \textit{Planck} CMB measurements and Sloan-Digital-Sky-Survey Lyman-$\alpha$-forest measurements \citep{gluscevic2017, boddy2018, xu2018, 2018arXiv180309734S}.
Upcoming measurements from Stage-3 experiments  (which we refer to as CMB-S3 in the following) such as AdvACT \citep{act2016} and SPT-3G \citep{spt2014}, and future experiments such as the Simons Observatory \citep{SOwebsite} and CMB Stage-4 (CMB-S4) \citep{s42016}, will measure small-scale anisotropy with a substantially higher precision than \textit{Planck} \citep{s42016}, and will thus enable a substantial improvement in constraining DM interaction physics.  
Furthermore, in the regime where DM is strongly coupled to baryons (at high redshifts, for a velocity-independent interaction), the effect of scattering resembles increasing the inertia of the baryon fluid, which in turns reflects as a shift of all CMB acoustic peaks in both temperature and polarization power spectra, towards higher multipoles \citep{boddy2018,chen2002,sigurdson2004}.
Thus, even the improvement of CMB measurements at large and moderate angular scales can contribute to a better constraint on DM-baryon scattering, in particular by achieving better resolution of the acoustic peaks in polarization.

In this work, we forecast the sensitivity of upcoming and future CMB measurements to constrain the DM-baryon scattering cross section, for DM masses down to 15 keV; we concentrate on the case of velocity-independent spin-independent elastic scattering--the primary interaction considered by many low-energy DM experiments~\citep{gaitskell2004}.
In terms of the cross section limit, compared to current \textit{Planck} results in \citep{gluscevic2017, boddy2018}, we find a factor of $\sim$6 improvement with CMB-S3, a factor of $\sim$26 with CMB-S4, and a factor of $\sim$200 for a cosmic-variance limited experiment.
The improvement is largely driven by the CMB lensing power spectrum measurement: we find that the addition of lensing renders our projected upper limits 4 to 6 times lower compared to the analysis of the primary CMB anisotropy alone.

We further investigate the dependence of the sensitivity to DM-baryon interactions on experimental design, and find that maximizing chances for detecting early-time interactions between DM and baryons with CMB requires a wide survey with high-resolution measurements of both the primary CMB and the CMB lensing power spectrum.

Finally, we examine potential degeneracies between signatures of DM-baryon scattering and other putative new-physics signals sought by the current and planned CMB experiments.
In particular, we quantify prospects for disentangling effects of DM-baryon interactions from DM annihilation signals, the signals from neutrino mass, and the signals from the effective number of relativistic degrees of freedom, using CMB measurements alone.
Detailed projections for sensitivity of CMB-S3 and CMB-S4 to DM annihilations and active neutrinos are presented in other literature (see \cite{s42016}), and our forecasts are in good agreement with previous results. 
We find no degeneracy between DM annihilations and other physical effects considered in this study.
Furthermore, in the primary CMB, we find a moderate degeneracy between the effects of the neutrino mass, relativistic degrees of freedom, and DM-baryon scattering.
Since the measurement of the neutrino mass is one of the primary science targets for the next-generation CMB experiments \citep{s42016}, 
our findings indicate the need for confirmation of a tentative signal with observables beyond the primary CMB, in order to break this degeneracy.
For this reason, we particularly focus on forecasts including the CMB lensing power spectrum.
We find that lensing-potential power spectrum measurements at the level of CMB-S4 promise to break degeneracies between all new physical effects considered in this study.

In Section \ref{sec:physics}, we provide a brief overview of the relevant modifications to standard cosmology needed to capture the effects of DM scattering, neutrino mass, the effective number of relativistic degrees of freedom, and DM annihilation on CMB power spectra.
In Section \ref{sec:fisher}, we summarize the Fisher information matrix method that we use in all our forecasting exercises, with some technical details presented in the Appendix.
In Section \ref{sec:results}, we present our numerical results. 
In Section \ref{sec:conclusion}, we make concluding remarks and discuss the implications of our findings.

\section{Modifications to Cosmology} \label{sec:physics}

\begin{figure}
\subfloat{
\centerline{
  \includegraphics[clip,width=0.85\columnwidth]{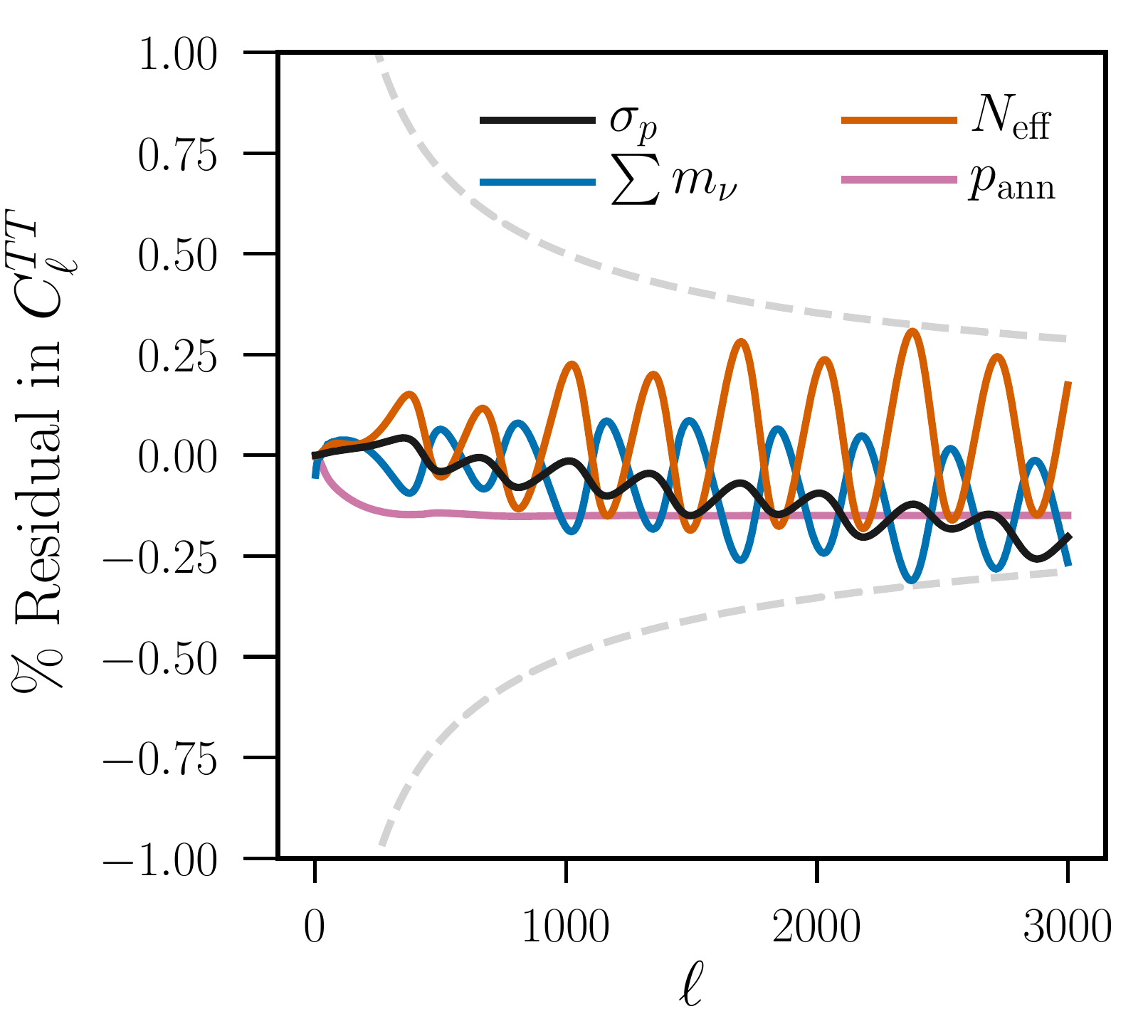}}%
}\\
\subfloat{
\centerline{
  \includegraphics[clip,width=0.85\columnwidth]{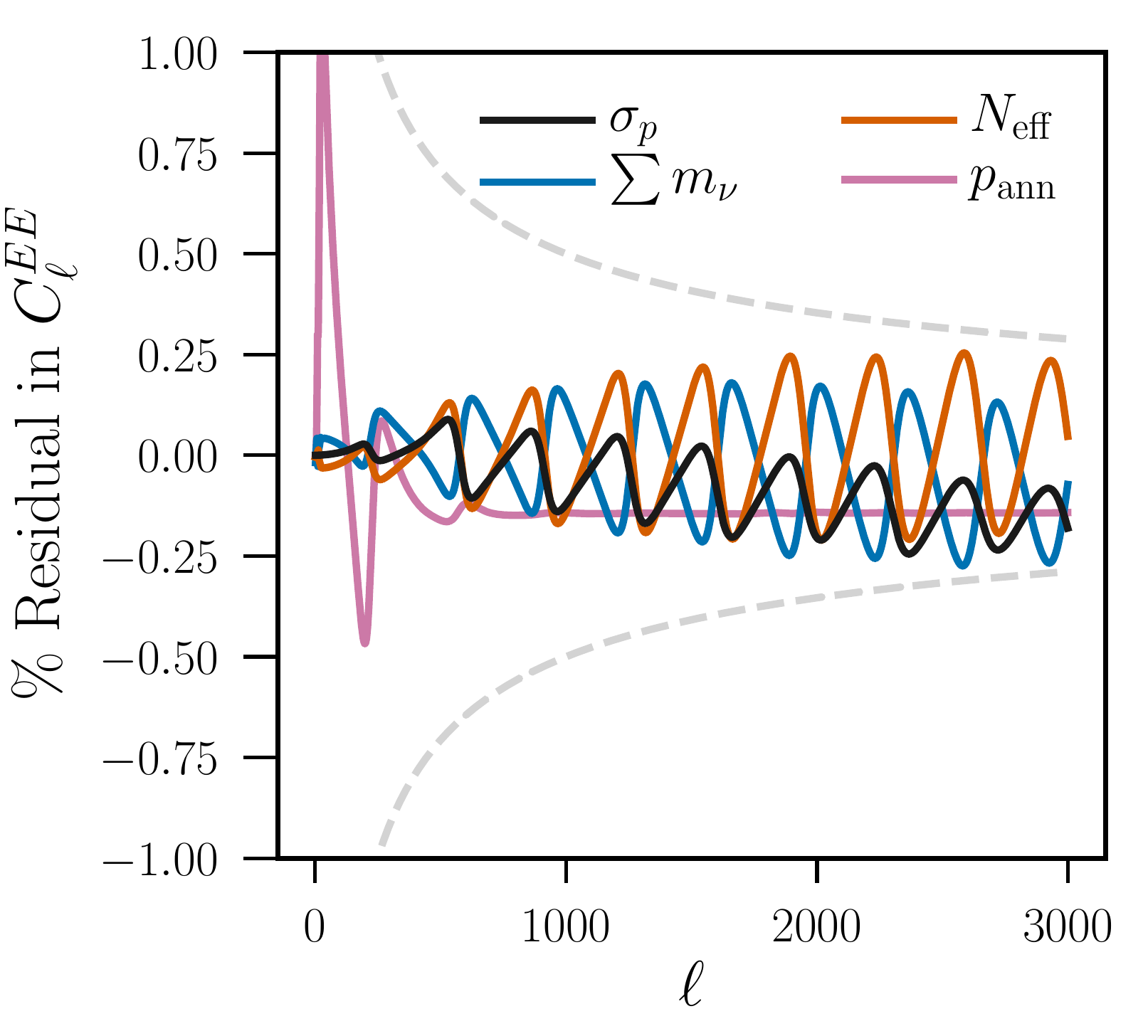}}%
}\\
\centerline{
\subfloat{
\centerline{
  \includegraphics[clip,width=0.85\columnwidth]{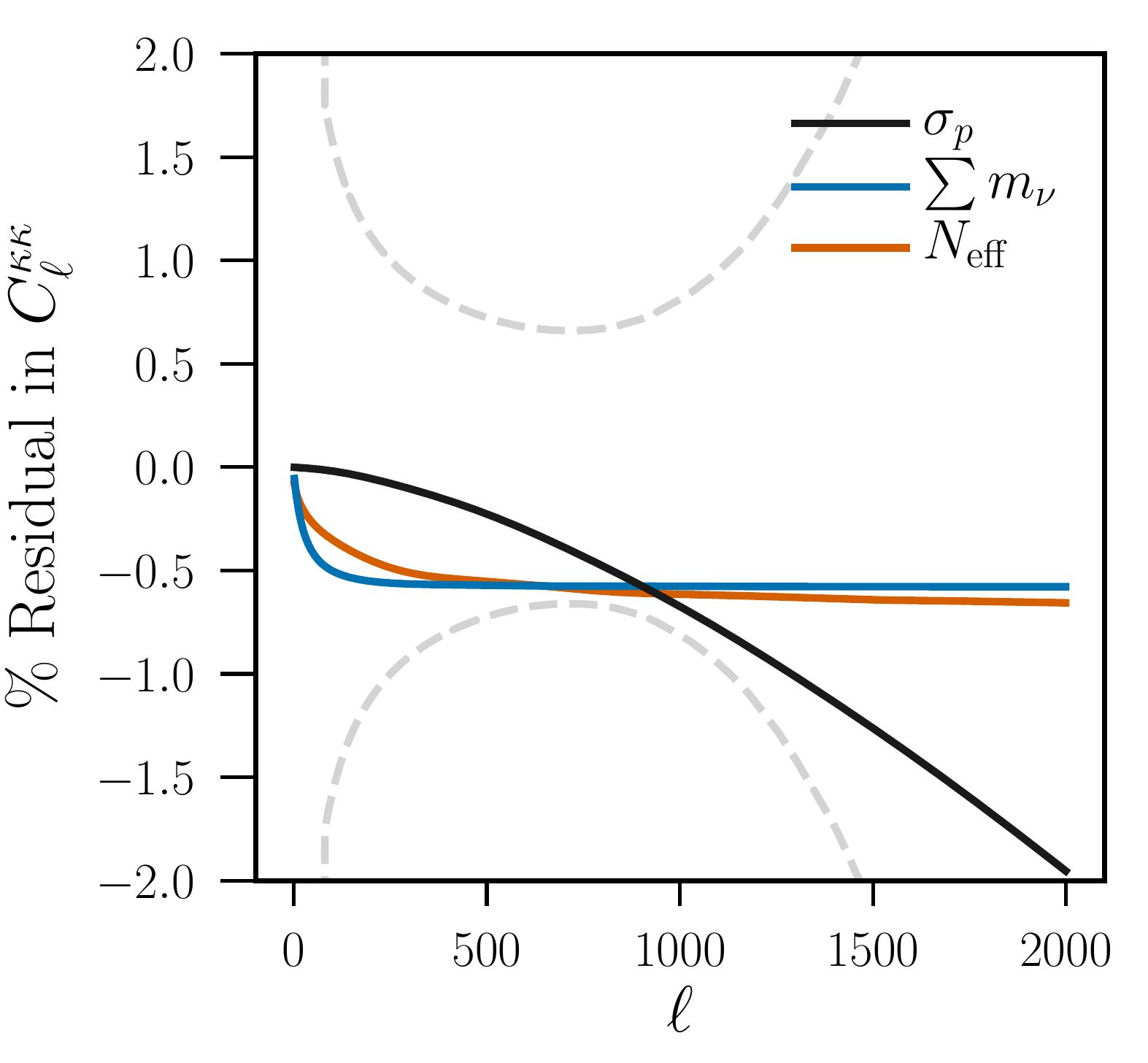}}%
}}
\caption{Fractional residuals of unlensed $C_{\ell}^{TT}$, $C_{\ell}^{EE}$, and $C_{\ell}^{\kappa \kappa}$.
Residuals are shown with respect to the $\Lambda$CDM power spectra, generated with a fiducial cosmology consistent with \textit{Planck}-2015 \citep{planck2015}. 
Modified cosmologies include---one at a time---the effects from non-vanishing: DM-proton scattering cross section $\sigma_p$, sum of the neutrino masses $\Sigma m_\nu$, non-standard effective number of relativistic degrees of freedom $N_\text{eff}$, and DM annihilation parameter $p_{\text{ann}}$. 
See text in Section \ref{sec:physics} for further details.
\label{fig:cartoon_cltteekk}}
\end{figure}

In this Section, we briefly review the basic physics behind DM-proton scattering, massive neutrinos, the effective number of relativistic degrees of freedom, and DM annihilation, and their effects on CMB power spectra. 
In Figure \ref{fig:cartoon_cltteekk}, we illustrate the effect of these non-standard cosmologies on the CMB temperature power spectrum $C_\ell^{TT}$, the $E$-mode polarization power spectrum $C_\ell^{EE}$, and CMB lensing convergence power spectrum $C_\ell^{\kappa\kappa}$ \citep{lewis2006}.
The lensing convergence power spectrum is computed with the linear matter power spectrum.
In this Figure, we show the fractional residual signal from the $\Lambda$CDM power spectra, varying each parameter individually while keeping all other parameters fixed. 
The values of the relevant parameters controlling the residual signal are set to their 2$\sigma$ upper limits derived using only the Fisher information from the power spectrum of interest using an experiment with CMB-S4 noise (as described in Section \ref{sec:results}), while fixing all other cosmological parameters to a set of values consistent with \textit{Planck}-2015 data (\cite{planck2015}, see Table \ref{tab:fiducial} in the Appendix for more details).
In the the top panel, we set: $\sigma_p=2.1\times 10^{-26}$ cm$^2$, $\sum m_{\nu} = 0.07$ eV, $N_{\text{eff}} = 3.06$, and $p_{\text{ann}} = 1.4 \times 10^{-7}$ m$^3$s$^{-1}$kg$^{-1}$; ii).
In the middle panel, we set: $\sigma_p=2.0\times 10^{-26}$ cm$^2$, $\sum m_{\nu} = 0.065$ eV, $N_{\text{eff}} = 3.05$, and $p_{\text{ann}} = 1.4 \times 10^{-7}$ m$^3$s$^{-1}$kg$^{-1}$; iii).
In the bottom panel, we set: $\sigma_p=8.8\times 10^{-27}$ cm$^2$, $\sum m_{\nu} = 0.08$ eV, and $N_{\text{eff}} = 3.096$.
The gray dashed curves represent the 1$\sigma$ error bars associated with a CMB-S4-like experiment with 1$\mu$K-arcmin noise, and $1.5$arcmin beam full-width-half maximum (FWHM), for the observed multipole range $2 < \ell < 3000$, binned in $\ell$ with a bin width of $\Delta\ell$=50.

\subsection{Dark Matter-Baryon Scattering} \label{subsec:DMeff}

We focus on spin-independent elastic scattering of DM with protons, with no dependence on relative particle velocity---a well-studied process described by a simple contact interaction, arising from, \textit{e.g.}, the exchange of a heavy scalar or vector mediator \citep{fan2010}.
Such a scenario is constrained both by \textit{Planck} measurements and by null results from direct detection experiments, for different ranges of DM masses.

Scattering of DM with baryons prior to recombination leads to momentum transfer and thus a drag force between the two fluids; this affects density and velocity perturbations, predominantly at small scales, but it also changes the global thermal history through modifications of fluid temperatures due to the friction force.

To compute the CMB power spectra in the non-standard scenario that accounts for DM-baryon scattering, we use the modified version of \texttt{CLASS} code \citep{class} developed for \cite{gluscevic2017}, with an implementation of the following modified Boltzmann equations 
\begin{equation}
\dot{\delta_{\chi}} = -\theta_{\chi} - \frac{\dot{h}}{2},
\end{equation}
\begin{equation}
\dot{\theta}_{\chi} = - \frac{\dot{a}}{a}\theta_{\chi} + c_{\chi}^2 k^2 \delta_{\chi} + R_{\chi} (\theta_b - \theta_{\chi}),
\end{equation}
for DM, and similarly for baryons,
\begin{equation}
\dot{\delta_{b}} = -\theta_{b} - \frac{\dot{h}}{2},
\end{equation}
\begin{equation}
\dot{\theta}_b = - \frac{\dot{a}}{a}\theta_b + c_b^2 k^2 \delta_b + R_{\gamma} (\theta_{\gamma} - \theta_{b}) + \frac{\rho_{\chi}}{\rho_b} R_{\chi} (\theta_{\chi} - \theta_b),
\end{equation}
where $\delta_{\chi}$ and $\delta_b$ are the density fluctuations in DM and baryons, respectively; $\theta_{\chi}$ and $\theta_b$ are the corresponding velocity divergences; $c_{\chi}$ and $c_b$ are the corresponding sound speeds; $\rho_{\chi}$ and $\rho_b$ are the corresponding energy densities;
$k$ is a wave number; $a$ is the scale factor; $h$ is the trace of the scalar metric perturbation; the overdot notation refers to a derivative with respect to conformal time; $\theta_{\gamma}$ refers to the velocity divergence for the photon component; $R_{\gamma}$ is a coefficient corresponding to the usual Compton scattering; and $R_{\chi}$ is the coefficient for the rate of momentum exchange between the DM and baryon fluids.
The strength of the interaction between DM and protons is parametrized by a coupling coefficient $c_p$ associated with an effective-theory operator describing the interaction \citep{fitzpatrick2013}. 
The coupling coefficient is related to the momentum-transfer cross section\footnote{For brevity, we refer to the momentum-transfer cross section as simply the cross section, as the two coincide for the interaction under consideration in this work.} as
$\sigma_{p}$=$\mu_{\chi p}^2/m_v^4 \pi c_p^2$, where $m_{v}$$\approx$246~GeV is the weak-scale mass (chosen as an arbitrary normalization), and $\mu_{\chi p}$ is the reduced mass of the DM-proton system.
We ignore scattering with neutrons, and include scattering on protons inside helium nuclei, following \citep{boddy2018}.
\footnote{This only affects our results for masses above a GeV, when scattering with helium becomes important.}
When averaged over the velocity distributions, the cross section leads to a momentum transfer coefficient; for scattering on proton, it is given by
\begin{equation}
R_{\chi p} = \mathcal{N}_0 a \rho_b (1 - Y_{\text{He}} ) \frac{ \sigma_p}{m_{\chi} + m_p} \left( \frac{T_b}{m_p} + \frac{T_{\chi}}{m_{\chi}} \right)^{1/2}, \label{eq:interaction_rate}
\end{equation}
where $\mathcal{N}_0 \equiv 2^{7/2}/3 \sqrt{\pi}$; $T_b$ and $T_{\chi}$ are the temperatures of the baryon and DM fluids, respectively; and $Y_{\text{He}}$ is the mass fraction in helium.
In the nuclear shell model, $a_{\text{He}} \approx 1.5$ fm parametrizes the size of the nucleus at hand.
The expression for the momentum rate coefficient for scattering on helium is somewhat more complicated, since it includes a nuclear form factor which yields velocity dependence even in our case of the interaction arising from a velocity-independent operator; we adopt the approach of \cite{boddy2018}, and refer the reader to it for further details on including scattering on helium.
In our case of spin-independent scattering, the total rate coefficient is the sum
over all species of nuclei that interact with DM---protons and helium nuclei.
The DM temperature is given as a solution to the following
\begin{equation}
\dot{T}_{\chi} = -2 \frac{\dot{a}}{a} T_{\chi} + 2 R_{\chi}' (T_b - T_{\chi}),
\end{equation}
where the heat-exchange coefficient $R_{\chi}'$ is closely related to the momentum-exchange coefficient, as detailed in \cite{boddy2018}.
We treat sub-GeV DM particles, so the $T_{\chi} / m_{\chi}$ term in Eq.~(\ref{eq:interaction_rate}) is non-zero, and we must solve for DM temperature evolution consistently. 
We ignore the back-reaction on the baryon temperature from scattering, which is negligible when baryons are tightly coupled to photons. 
Indeed, in the velocity-independent scenario considered in this work, the effect of interactions is dominant only before recombination, so the back-reaction term is negligible for all times where scattering is important.
We ignore the relative bulk velocity of the DM and baryon fluids, as it is negligible at relevant cosmological times \citep{dvorkin2014, gluscevic2017}.

In terms of the CMB observables, the primary effect of DM-baryon scattering is a suppression of power on small angular scales (see Figure \ref{fig:cartoon_cltteekk}).
In addition, scattering changes the baryon speed of sound, shifting the angular scale of the acoustic peaks in the CMB power spectra.
Since the peaks are narrower in multipole space in the case of the polarization power spectra, we may expect that polarization can substantially lower the upper limits on DM-baryon interaction strength.
However, for \textit{Planck}'s noise levels, polarization and CMB lensing power spectrum measurements improve the current constraints by only $\sim$30\% \citep{gluscevic2017, boddy2018}.
As we show in Section \ref{subsec:experimentaldesign}, the contribution of lensing and polarization to constraining DM-baryon interactions will substantially increase with next-generation experiments.

Finally, scattering of DM with baryons can force the DM fluid to undergo acoustic oscillations if coupling is strong at early times, and this produces oscillatory features and suppression of power on small scales in the linear matter power spectrum~\citep{boddy2018}; this effect too can be used to search for evidence of DM interactions with large-scale-structure observables~\citep{chen2002, sigurdson2004, dvorkin2014}.
In addition, modifications to the matter power spectrum imprint on secondary CMB anisotropies, and can be captured using the power spectrum of the convergence $\kappa$ in the lensing field, reconstructed as a 4-point function from CMB maps \citep{hu2002}. 
In Section \ref{sec:results}, we examine sensitivity of both 2-point and 4-point functions to capturing the effects of DM-baryon scattering. 
The effect of DM-proton scattering on temperature, polarization, and the CMB lensing power spectrum is illustrated in Figure \ref{fig:cartoon_cltteekk}.

\subsection{Neutrino Mass}

The sum of the masses of the three neutrino species, $\Sigma m_\nu$, is currently not known, but the differences in the squared masses have been measured through oscillation experiments \citep{patrignani2016}.
These mass differences allow for two ways to order the masses of the neutrino species: a ``normal" hierarchy where the sum of the masses is dominated by one species, and an ``inverted" hierarchy where two neutrino species are much more massive than the third. 
In this work, we assume the normal mass hierarchy, featuring a squared mass difference of $2 \times 10^{-3}$ eV$^2$ between two neutrino species, and a negligible mass difference between the two lower-mass species. 

Upon decoupling from the thermal bath (at temperatures of about $1$ MeV), massive neutrinos are initially relativistic. 
As the Universe expands, they eventually become nonrelativistic and cosmologically indistinguishable from matter \citep{lesgourgues2006, allison2015}; the transition happens at a redshift $z_\text{nr}$$\approx$$120\times\left( m_\nu/60\text{ meV}\right)$ \citep{ichikawa2005}.
However, unlike cold DM (CDM), once they become nonrelativistic at a redshift of a few hundred, they still have a finite velocity dispersion (and thus behave like \emph{warm} DM). 
Thus, they do not cluster like CDM and baryons, on scales smaller than their free-streaming length ($1/k_{\mathrm{fs}}$, the typical distance a neutrino travels in a Hubble time before scattering).
For modes $k\gg k_{\mathrm{fs}}$, they do not contribute to the fluctuations in the source potentials, but they still contribute to the Hubble expansion rate through their energy density. 
Hence, neutrino mass affects both the expansion history and the evolution of perturbations: the matter fluctuations grow slower in presence of neutrino mass, resulting in a scale-dependent suppression of gravitational potentials and of the matter power spectrum, relative to a cosmology with massless neutrinos.

The effects of non-vanishing neutrino mass are imprinted on both the primary CMB anisotropy and in the amount of gravitational lensing of the primary CMB. 
Indeed, as photons leave the last-scattering surface at $z=1100$, they undergo deflections as they pass through growing gravitational potentials of matter fluctuations; in a cosmology with massive neutrinos, suppressed matter fluctuations lead to a reduction in the amount of lensing the CMB experiences. 
At the level of the temperature and polarization power spectra, lensing appears primarily as a smearing of the acoustic peaks, and an increase in power at high $\ell$ multipoles.
The presence of massive neutrinos shows up as a reduction in the amount of peak smearing and a reduction of power at small angular scales, in the 2-point functions ($TT$, $EE$, and $TE$), as compared to the $\Lambda$CDM case.
Additionally, massive neutrinos are more directly probed through the lensing convergence power spectrum $C_\ell^{\kappa\kappa}$, reconstructed from CMB maps as a 4-point function \citep{hu2002}.
To compute the effect of $\Sigma m_\nu$ on the CMB power spectra, we use the existing implementation of ``non-cold dark matter relics'' in \texttt{CLASS}; we illustrate these effects in Figure \ref{fig:cartoon_cltteekk}.

\subsection{Effective number of relativistic species}

If there are additional relativistic degrees of freedom in the early Universe, their contribution to the radiation energy density $\rho_\text{rad}$ can be parameterized by the effective number of relativistic species $N_\text{eff}$ as
\begin{equation}
  \rho_\text{rad} = \left[1+N_\textrm{eff} \frac{7}{8} \left(\frac{4}{11}\right)^{4/3} \right]\rho_\gamma ,
\end{equation}
where $\rho_\gamma$ represents the energy density of the CMB photons. 
In the Standard Model, $N_\textrm{eff}$=$3.046$ accounts for the three massless active neutrinos; a departure from this value would imply presence of new relativistic particles.
Such presence would impact both the overall expansion history and the evolution of perturbations; the dominant effect would arise from increasing the expansion rate around the time of recombination, which would shift the acoustic peaks to larger angular scales, and suppress power on scales that entered the sound horizon prior to recombination \citep{bashinsky2004, hou2013, lesgourgues2013, baumann2017}. 
In terms of the CMB lensing power spectrum, the main effect comes through the suppression of power that results from shifting matter-radiation equality to later times.
In this work, we assume the new relativistic species is free-streaming, and utilize the implementation of $N_\text{eff}$ already present in the \texttt{CLASS} code; the effects of $N_\text{eff}$ on the CMB power spectra are illustrated in Figure \ref{fig:cartoon_cltteekk}.

\subsection{Dark Matter Annihilation}

If DM particles annihilate into Standard Model particles, they can modify the recombination history of the Universe. 
Namely, annihilation injects energy into the pre-recombination photon-baryon plasma and post-recombination gas and background photons, which can leave an imprint in the CMB power spectra.
Indeed, large rates of energy injection are ruled out by observations of the positions of the first few peaks
of the CMB temperature power spectrum \citep{planck2015}. 
For much lower rates of energy injection, the primary
effect on the CMB is due to an effective broadening of the surface of last scattering, which leads to a suppression of fluctuations at small angular scales. 
Energy injection from annihilation also shifts the acoustic peaks in $TE$ and $EE$, and enhances $EE$ power at $\ell$$<$$500$, by effectively increasing the thickness of the last scattering surface \citep{galli2009, slatyer2009, finkbeiner2012, madhavacheril2014, green2018}.
While constraints from $TT$ are not expected to improve much with future CMB experiments, there is considerable room for improvement through more precise measurements of polarization \citep{madhavacheril2014}.

Previous studies \citep{finkbeiner2012,madhavacheril2014} have shown that the effects of an arbitrary annihilation history on the CMB are well captured by a single parameter $p_\text{ann}$ corresponding to the amplitude of the first principal component obtained in \cite{finkbeiner2012}.
We parametrize the DM annihilation signal with $p_{\text{ann}}$, such that the rate of energy deposition per unit volume and time is given by
\begin{equation}
\left( \frac{dE}{dt \, dV} \right)_{\text{ann}} = p_{\text{ann}}(z) c^2 \Omega_\text{DM}^2 \rho_c^2 (1+z)^6, \label{eq:ann}
\end{equation}
where $p_{\text{ann}}(z)$$=$$p_{\text{ann}} f(z)$, and $f(z)$ represents the redshift-dependent first principal component; $\Omega_\text{DM}$ is the energy density in DM.
The \texttt{CLASS} code already implements DM annihilations in the thermodynamics module, which integrates the ionization and matter temperature in the case where $p_{\mathrm{ann}}(z)$ is constant \citep{giesen2012}. 
In order to apply the ``universal'' redshift dependence, we modify \texttt{CLASS} to include $f(z)$ from \cite{finkbeiner2012} in combination with the on-the-spot approximation, in which we assume that DM annihilation deposits energy instantaneously into the local plasma. 
The resulting effect of DM annihilations on the CMB power spectra is illustrated in Figure \ref{fig:cartoon_cltteekk}.
We note that the annihilation signal makes virtually no impact on the lensing convergence signal, so we omit the corresponding line from the bottom panel of this Figure.

\section{Forecasting Method}\label{sec:fisher}

We now describe our method for forecasting the sensitivity of various CMB experiments to detecting physical effects described in the previous Section, and we summarize experimental parameters and assumptions used in all our numerical analyses.

\subsection{Fisher Formalism}

We use Fisher formalism and compute Fisher-information matrices whose components are given by \citep{eisenstein1999,coe2009,wu2014}
\begin{equation}
F_{ij} = \sum_\ell \frac{2\ell+1}{2} f_\mathrm{sky} \text{Tr}\left( \mathcal{C}_{\ell}^{-1}(\vec{\theta}) \frac{\partial \mathbf{C}_{\ell}}{\partial \theta_i} \mathcal{C}_{\ell}^{-1}(\vec{\theta}) \frac{\partial \mathbf{C}_{\ell}}{\partial \theta_j}  \right), \label{eq:fisher}
\end{equation}
where $f_\mathrm{sky}$ is fractional sky coverage; Tr stands for the trace of a matrix; $\vec \theta$ represents the vector of model parameters, which we fit for when analyzing data; $C_\ell$'s represent our data set; and the covariances are given by
\begin{equation}
\mathbf{C}_{\ell} \equiv \left(
\begin{array}{ccc}
 C_{\ell}^{TT} + N_{\ell}^{TT} & C_{\ell}^{TE} & 0 \\
 C_{\ell}^{TE}                 & C_{\ell}^{EE} + N_{\ell}^{EE}   & 0 \\
 0 & 0 & C_{\ell}^{\kappa\kappa} + N_{\ell}^{\kappa \kappa} \\
\end{array}
\right).
\end{equation}
We assume a white-noise power spectrum $N_\ell^{XX'}$ for effective noise, where $XX'\in\{TT,EE,TE\}$, is given by
\begin{equation}
N_\ell^{XX'} = s^2 \text{exp} \left( \ell(\ell+1) \frac{\theta^2_\mathrm{FWHM}}{8 \ln 2} \right),
\end{equation}
the resolution of the experiment in arcmin is $\theta_\mathrm{FWHM}$; $s$ is the instrumental noise in temperature; and $\sqrt{2}s$ the noise in polarization. 
We neglect the off-diagonal $T\kappa$ and $E\kappa$ terms, which have very little contribution to our constraints \citep{wu2014}. 

\begin{figure}
\centerline{
\epsscale{1.1}
\plotone{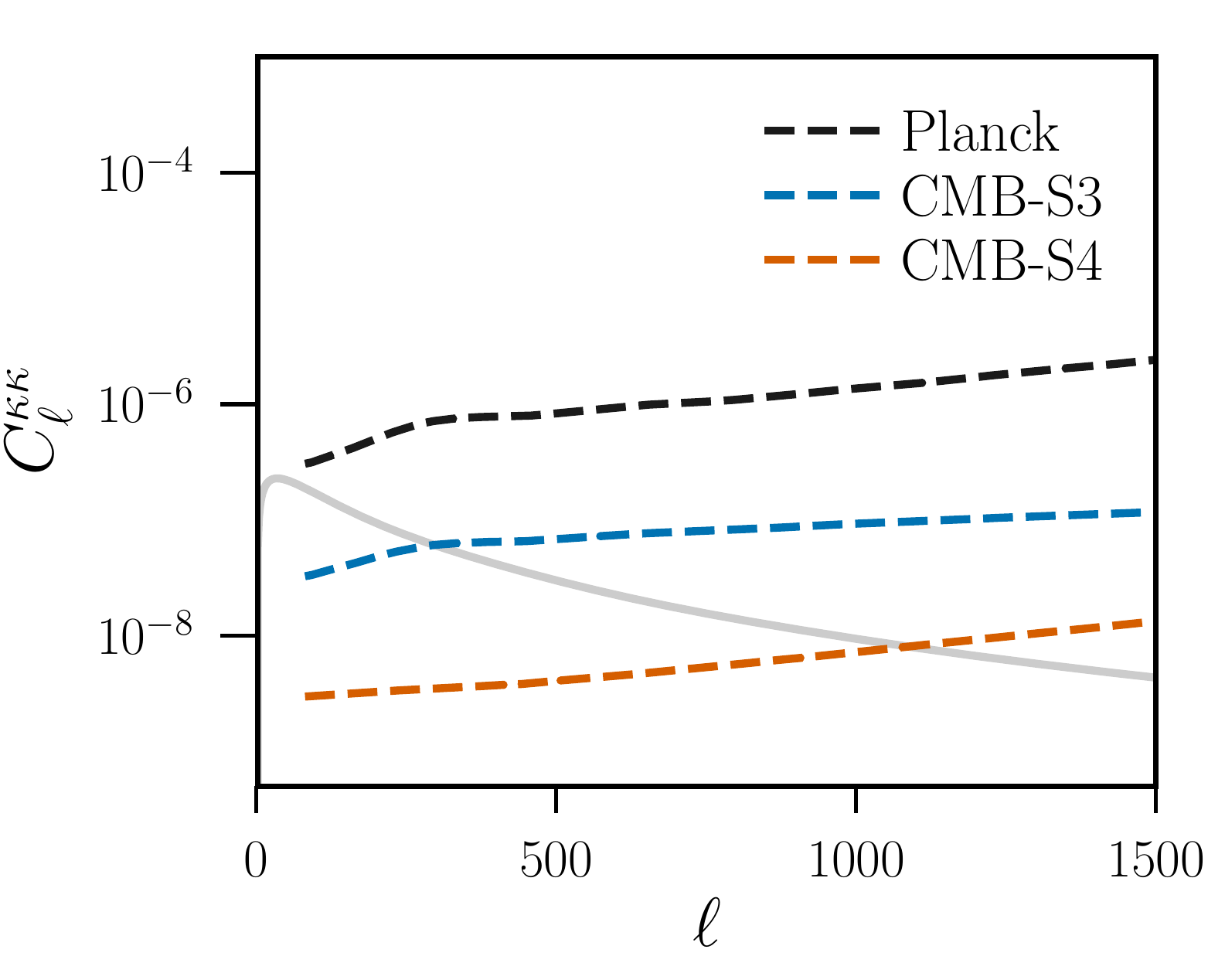}}
\caption{The CMB lensing convergence noise spectrum $N_{\ell}^{\kappa \kappa}$ is shown for \textit{Planck}, CMB-S3, and CMB-S4 experiments (from top to bottom colored curve on the right-hand side of the plot, respectively, all dashed lines), compared to the lensing convergence signal $C_{\ell}^{\kappa \kappa}$ (solid), computed with parameters consistent with the \textit{Planck}-2015 best-fit cosmology \cite{planck2015} (see Table \ref{tab:fiducial} in the Appendix). 
For different experiments, we use the noise levels in Table \ref{tab:experiments}. \label{fig:lensing_noise}}
\end{figure}

For the lensing convergence, we use the \texttt{orphics} code \citep{madhavacheril2018} to calculate the noise spectrum $N_{\ell}^{\kappa \kappa}$ from a minimum variance combination of $TT$, $TE$, $EE$, $EB$ and $TB$ quadratic estimators, assuming the CMB B-modes undergo an iterative delensing procedure. 
For all experiments, we compute $C_{\ell}^{\kappa \kappa}$ over 20$<$$\ell$$<$2500, and assume infinite noise outside this multipole range. 
We compute the Planck lensing noise using temperature and polarization multipoles 2$<$$\ell$$<$3000.
For CMB-S3 and CMB-S4, we compute with temperature multipoles 300$<$$\ell$$<$3000 and polarization multipoles 300$<$$\ell$$<$5000 \citep{s42016}.
Our cosmic variance limited experiment is assumed to have zero noise in $\kappa\kappa$ over the entire reconstructed multipole range of 20$<$$\ell$$<$2500.
We show the lensing noise estimates in Figure \ref{fig:lensing_noise}.
We cross-validated our noise calculation for the CMB-S4 case against the estimates in \citep{s42016}. 

To avoid double-counting lensing information in both $C_{\ell}^{\kappa\kappa}$ and in the CMB primary anisotropy, we use the unlensed $C_{\ell}^{TT}$, $C_{\ell}^{TE}$, $C_{\ell}^{EE}$ spectra when they are analyzed in combination with $C_{\ell}^{\kappa \kappa}$. 
A more precise joint treatment of both the lensed spectra and convergence is possible, but requires analytic calculation or joint simulation of CMB 2-point and lensing reconstruction power spectra to estimate the joint covariance matrix \citep{benoitlevy2012,schmittfull2013,peloton2017}. 
In ignoring the lensing information from the CMB primary anisotropy, we have thus made a conservative forecasting choice.

Furthermore, for the purposes of our Fisher analyses, we do not model nonlinearities in the matter power spectrum when computing the CMB lensing signal; this too is a conservative choice. 
Namely, the nonlinear power spectrum has a higher small-scale lensing signal in $\Lambda$CDM, so the fractional suppression in the small scale matter power from various effects considered in this work is more evident in a nonlinear treatment.
However, fitting function methods for handling nonlinear evolution like \texttt{Halofit} \citep{smith2003, bird2012} are not calibrated for cosmologies that include neutrinos and DM-baryon scattering, and we do not use them in this work.

To evaluate Fisher matrices, we developed a modified Fisher-analysis approach, which supplements the standard Fisher calculation with a covariance-matrix sampling procedure; our method is discussed in detail in the Appendix.
Its purpose is to provide a higher accuracy compared to the standard Fisher method, when the posterior probabilities are non-Gaussian (like in the case of $\sigma_p$, as reported in previous data analyses), and to enable implementation of arbitrary priors on chosen parameters, while preserving computational speed on a par with that of the stadard Fisher analyses.  
We choose flat priors for all parameters of interest in this study.\footnote{We note, however, that the choice of prior may produce order-unity variation in the forecasted upper bounds \citep{hannestad2017}.}, and we verify the validity of our forecasts by comparing our results with the forecasts obtained by MCMC methods on simulated data, as discussed in detail in the Appendix.

\subsection{Experimental Parameters}

\begin{deluxetable*}{c|c|c|c|c}
\tablehead{
\colhead{Experiment} & \colhead{$\ell$-range} & \colhead{Noise $s$} & \colhead{$f_\mathrm{sky}$} & \colhead{$\theta_\mathrm{FWHM}$}  \\
\colhead{} & \colhead{} & \colhead{[$\mu$K-arcmin]}  & \colhead{}  & \colhead{[arcmin]}
}
\startdata
\begin{tabular}{@{}c@{}}\textit{Planck} \\ (standalone)\end{tabular} &  \begin{tabular}{@{}c@{}}T: 2-2500 \\ P: 30-2500\end{tabular} & \begin{tabular}{@{}c@{}}T: 145,149,137,65,43,66,200 \\ P: -,-,450,103,81,134,406 \end{tabular} & \begin{tabular}{@{}c@{}} 0.6 \end{tabular} & 33,23,14,10,7,5,5 \\ \hline
\begin{tabular}{@{}c@{}}\textit{Planck} \\ (with S3/S4)\end{tabular} &  \begin{tabular}{@{}c@{}} T: 2-100 \\ P: 30-100 \\ overlap: 100-2500 \end{tabular} & $\cdots$ & \begin{tabular}{@{}c@{}}  0.6 \\ overlap: 0.2 \end{tabular} & $\cdots$ \\ \hline
CMB-S3  &  T/P: 100-3000 &10 & 0.4 & 1.4 \\ \hline
CMB-S4  &  \begin{tabular}{@{}c@{}} T: 300-3000 \\ P: 100-5000 \end{tabular} & \begin{tabular}{@{}c@{}}1.0 \end{tabular} & 0.4 & 1.5 \\ \hline
CV-limited  &  \begin{tabular}{@{}c@{}} T/P: 2-5000 \end{tabular} & 0.0 & 1.0 & N/A \\
\enddata
\caption{The experimental parameters used for our Fisher analyses. 
Parameters are listed for CMB-S3 (S3) and CMB-S4 (S4) experiments, and for a mock-version of \textit{Planck} \citep{allison2015, s42016,madhavacheril2017}.
When \textit{Planck}'s parameters are chosen depending on whether it is considered as a standalone experiment, or in combination with S3/S4, to avoid double-counting of modes.
Parameters denoted as ``overlap'' are used for the fraction of the sky that is observed by both \textit{Planck} and the S3/S4 experiment; low multipoles are always analyzed assuming $f_\text{sky}=0.6$. 
``T'' denotes parameters used for temperature-only Fisher analysis, and ``P'' corresponds to the parameters used when polarization is considered.
We specify the noise and beam widths $\theta_\mathrm{FWHM}$ for the individual \textit{Planck} frequency bands, combined with inverse-variance weights. 
}
 \label{tab:experiments}
\end{deluxetable*}

The noise properties of each of the experiments we consider in this work are listed in Table \ref{tab:experiments}.
When estimating \textit{Planck} noise levels, we follow \citep{allison2015}, which includes effective $TT$ noise matching \textit{Planck}-2015 data release, and $TE$ and $EE$ noise at their Bluebook values \citep{bluebook2006}. 
We also consider a generic CMB-S3 experiment, choosing noise parameters to be representative of the reach of the near-future CMB measurements with AdvACT \citep{act2016} and SPT-3G \citep{spt2014}, featuring a large improvement of sensitivity at small angular scales \citep{madhavacheril2017}.
Finally, we consider one possible configuration of the planned CMB-S4 experiment, which should improve upon CMB-S3 roughly by an order of magnitude in effective noise levels, according to the community planning report of \cite{s42016}.

Being ground-based observatories, CMB-S3 and CMB-S4 will have difficulty capturing the largest angular scales, and will also feature a limited sky coverage.
We thus always consider their measurements in combination with \textit{Planck} data, especially when analyzing power spectra at $\ell$$<$$100$. 
To avoid double-counting when performing combined analyses, we use $f_\text{sky}$ for \textit{Planck} that accounts only for the areas of the sky not covered by the ground-based experiments; this adjustment is described in the second row of Table \ref{tab:experiments}. 
We also produce constraints for a standalone, full-sky cosmic-variance limited (CV-limited) experiment, in which we set the noise in temperature, polarization, and lensing convergence to zero for $2 < \ell < 5000$.

\section{Results}\label{sec:results}

We now present projected constraints for the velocity-independent spin-independent DM-proton scattering, and review forecasts for other physical effects: effective number of relativistic degrees of freedom, sum of the neutrino masses, and DM annihilations, for upcoming and future CMB experiments.
In particular, for CMB-S4, we explore the impact of experimental design and survey strategy when forecasting sensitivity to DM interactions.
At the end, we quantify degeneracies between the various signatures of physics beyond the Standard Model.

\subsection{Forecasts for DM-baryon scattering}

\begin{figure}
\centerline{
\plotone{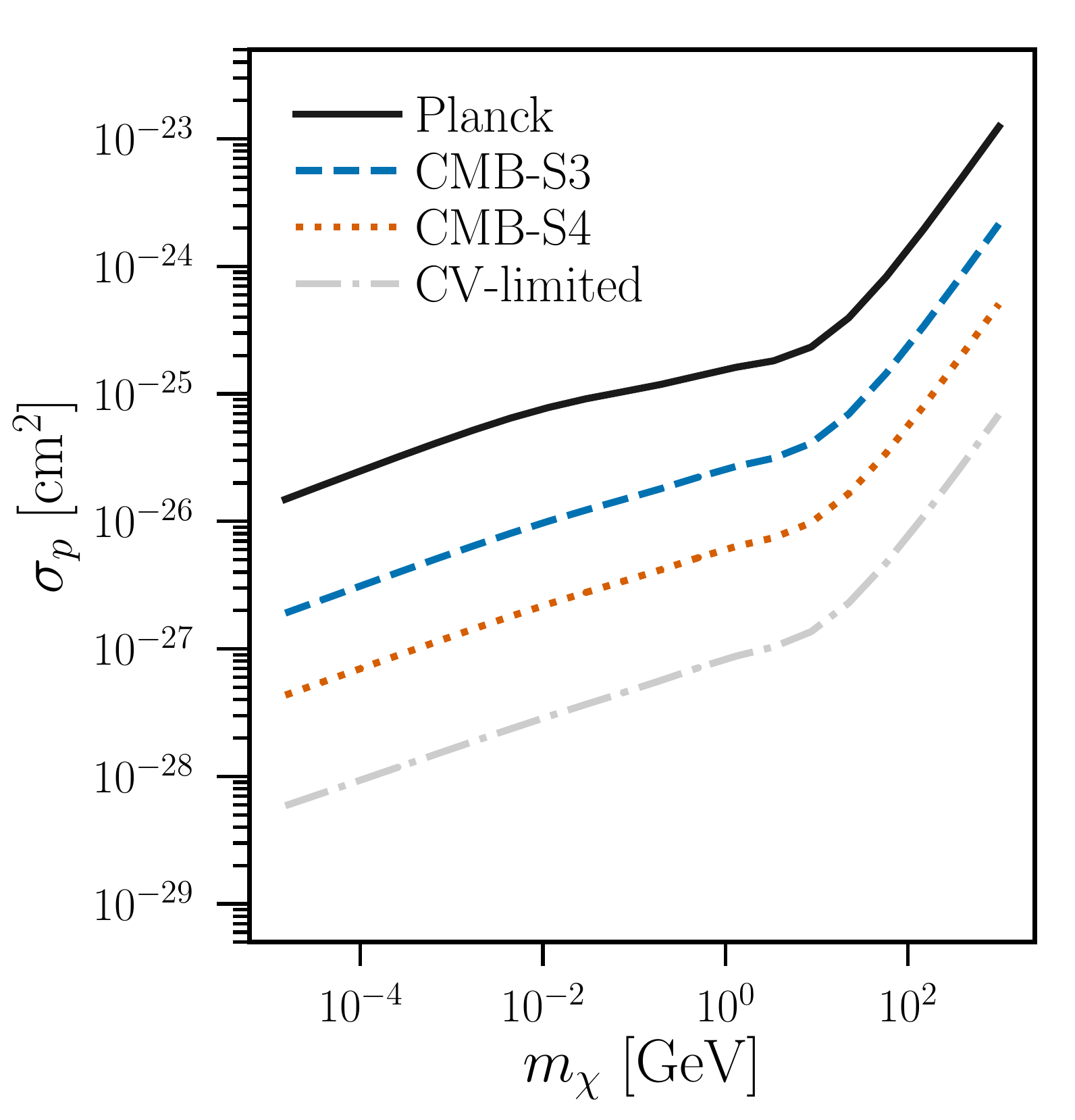}}
\caption{Projected 2$\sigma$ upper limits for the DM-proton scattering cross section for \textit{Planck}, CMB-S3, CMB-S4, and cosmic variance limited experiments, obtained assuming a null signal. 
Experimental parameters used in Fisher analyses are listed in Table \ref{tab:experiments}.\label{threeinstrument}}
\end{figure}

We start by simultaneously projecting constraints on the standard cosmological parameters (baryon density $\Omega_bh^2$, DM density $\Omega_\mathrm{DM} h^2$, the Hubble parameter $h$, the reionization optical depth $\tau_\mathrm{reio}$, the amplitude of the scalar perturbations $A_s$, and the scalar spectral index $n_s$), along with the coupling constant $c_p$ for velocity-independent spin-independent DM-proton scattering.
We consider three experiments: a mock version of \textit{Planck}, CMB-S3, and CMB-S4.
We analyze CMB primary temperature and polarization anisotropy captured by the unlensed $TT$, $EE$, and $TE$ power spectra, in conjunction with the lensing convergence power spectrum.
As our fiducial model, we choose the null-case best-fit \textit{Planck}-2015 $\Lambda$CDM cosmology, with three massless active neutrinos, no DM-baryon interactions and annihilations, and no new relativistic particles.
For a study case of 1 GeV DM particle, we obtain the following 2$\sigma$ upper limits on the coupling coefficient:
\begin{equation}
c_p <
\left\{
    \begin{array}{ll}
        4.4 \times 10^6  & \mbox{\quad(\textit{Planck})}  \\
        1.8 \times 10^6 & \mbox{\quad(CMB-S3)}  \\
        0.9 \times 10^6 & \mbox{\quad(CMB-S4)} \\
        0.3 \times 10^6 & \mbox{\quad(CV-limited)}
    \end{array}
\right.
\end{equation}
and in terms of the scattering cross section:
\begin{equation}
\frac{\sigma_p}{\text{cm}^2} <
\left\{
    \begin{array}{ll}
        16 \times 10^{-26}  & \mbox{\quad(\textit{Planck})}  \\
        2.6 \times 10^{-26} & \mbox{\quad(CMB-S3)}  \\
        0.6 \times 10^{-26} & \mbox{\quad(CMB-S4)} \\
        0.08 \times 10^{-26} & \mbox{\quad(CV-limited)} 
    \end{array}
\right.
\end{equation}
The main result of our forecasting exercises for DM-proton interactions is presented in Figure \ref{threeinstrument}, which shows the projected 2$\sigma$ exclusion curves in the cross section--DM-mass parameter space, for DM masses between 15 keV and 1 TeV.
Our \textit{Planck} projections are in good agreement with results from \textit{Planck} data analyses done in previous studies \citep{gluscevic2017, boddy2018}.

\begin{figure*}
\centerline{
\epsscale{1.1}
\plotone{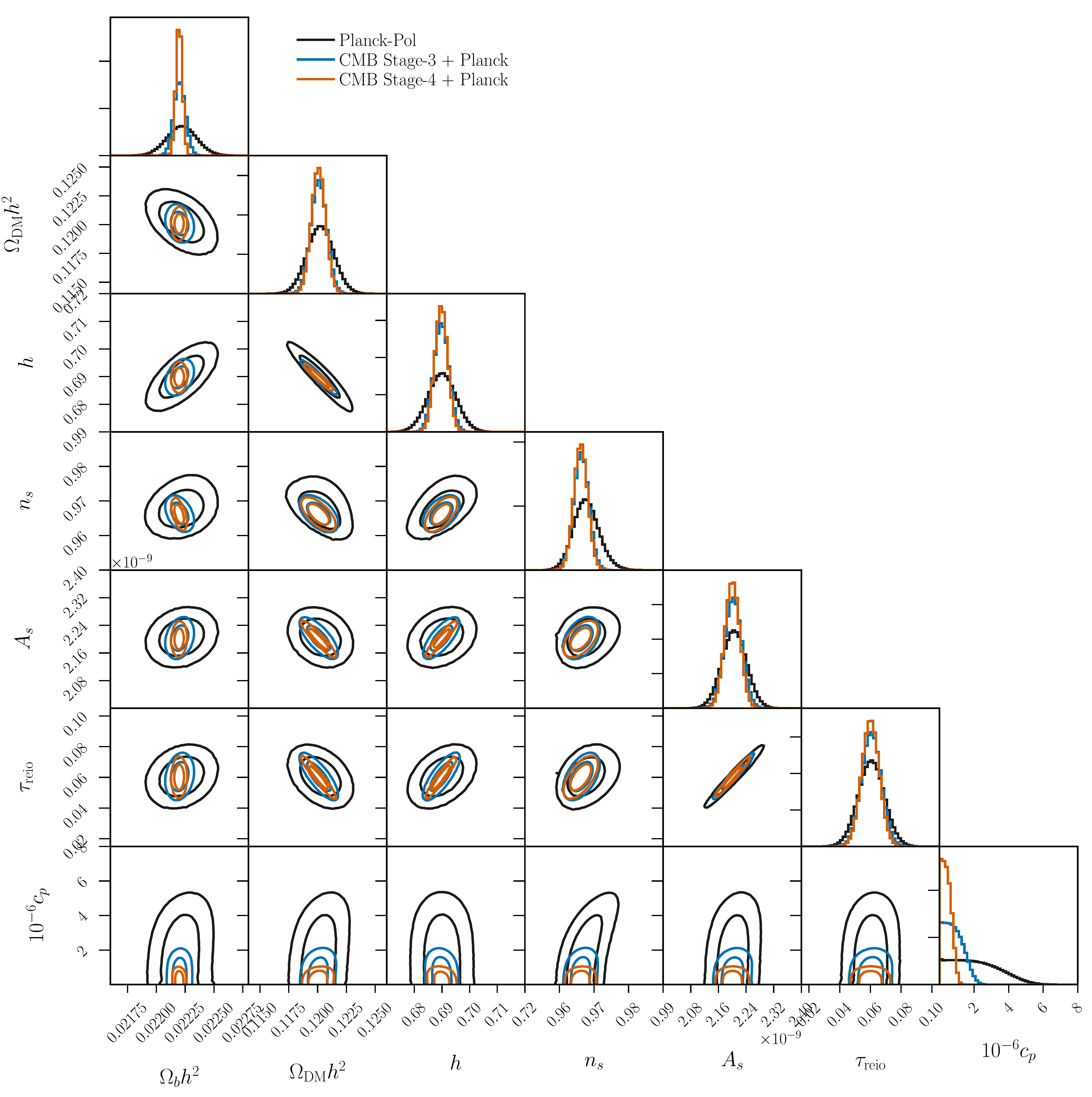}}
\caption{Fisher forecasts for the $\Lambda$CDM parameters and the coupling coefficient for velocity-independent spin-independent DM-proton scattering, for \textit{Planck}, CMB-S3, and CMB-S4 experiments. 
Shown are the 1$\sigma$ and 2$\sigma$ error ellipses (and the corresponding marginalized 1d posterior probability distributions on the top of each row), computed assuming the best-fit \textit{Planck}-2015 cosmology with no DM-proton interactions (zero coupling) as the fiducial model.
Experimental parameters used in Fisher analyses are listed in Table \ref{tab:experiments}.
\label{full_triangle_cc}}
\end{figure*}

Figure \ref{full_triangle_cc} shows the projected 1$\sigma$ and 2$\sigma$ error ellipses for the standard cosmological parameters and the DM-proton coupling constant, for the three CMB experiments, assuming a fixed DM particle mass of 1 GeV.
From Figure \ref{full_triangle_cc}, we see that, for \textit{Planck}, the coupling strength $c_p$ is most degenerate with the scalar spectral index $n_s$: the increase in the value of $c_p$ leads to a larger drag force between DM and baryons, producing a more prominent suppression of power at small angular scales; an increase in the value of $n_s$ can partly compensate for the suppression by tilting the spectrum.
However, the spectral tilt affects all scales, while the momentum exchange between DM and baryons produces a scale-dependent effect. 
The degeneracy between $n_s$ and $c_p$ is thus substantially reduced with better anisotropy measurements at intermediate scales in the case of CMB-S3 experiment, as evident in this Figure. 

The factor of $\sim$6 improvement in $\sigma_p$ constraints with
CMB-S3, as compared to Planck, comes from a combination of improvements in the overall $TT$ noise, as well as from new small-scale temperature information from 2500$<$$\ell$$<$3000. 
The additional factor of $\sim$4 between CMB-S3 and CMB-S4 comes from further improvements in small-scale anisotropy measurements,
especially in lensing. 
To illustrate where the information on scattering is coming from in the power spectra, we plot the summands of Eq.~(\ref{eq:fisher}), for \textit{Planck} (top panel) and CMB-S4 (bottom panel).
The diagonal elements of the Fisher information matrix shown in this Figure correspond to the inverse covariance of the parameter $\sigma_p$ (ignoring correlations between parameters), and thus serve as a proxy for constraining power on that parameter in a given experiment.
For the case of \textit{Planck}, Fisher information is dominated by temperature at all angular scales; contribution from polarization is only significant at $\ell$$\lesssim$1000, while contribution from lensing is negligible. 
However, as we discuss in Section \ref{subsec:degeneracy}, the lensing information reduces degeneracies between DM scattering and other parameters. This changes the non-diagonal terms in the Fisher matrix, and leads to a moderate impact on the marginalized distribution of $\sigma_p$.

For the case of CMB-S4, temperature and polarization make similar contributions to the constraint; however, lensing becomes the
dominant source of information on all angular scales (see
the bottom panel of the same Figure).
The situation is similar for the cosmic variance-limited experiment, with information primarily coming from the small scale lensing power spectrum.

\begin{figure}
\centerline{
\plotone{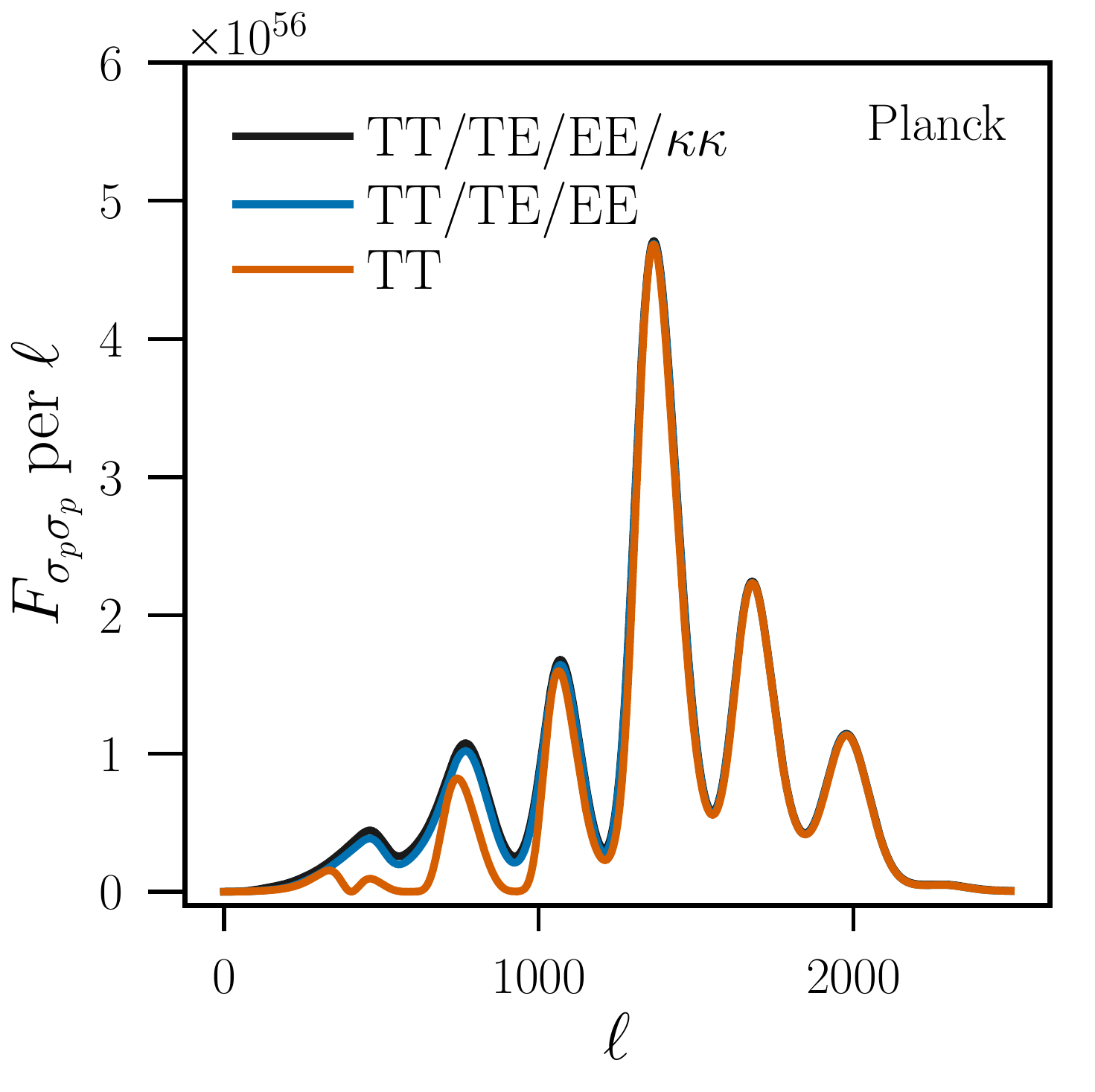}}
\centerline{
\plotone{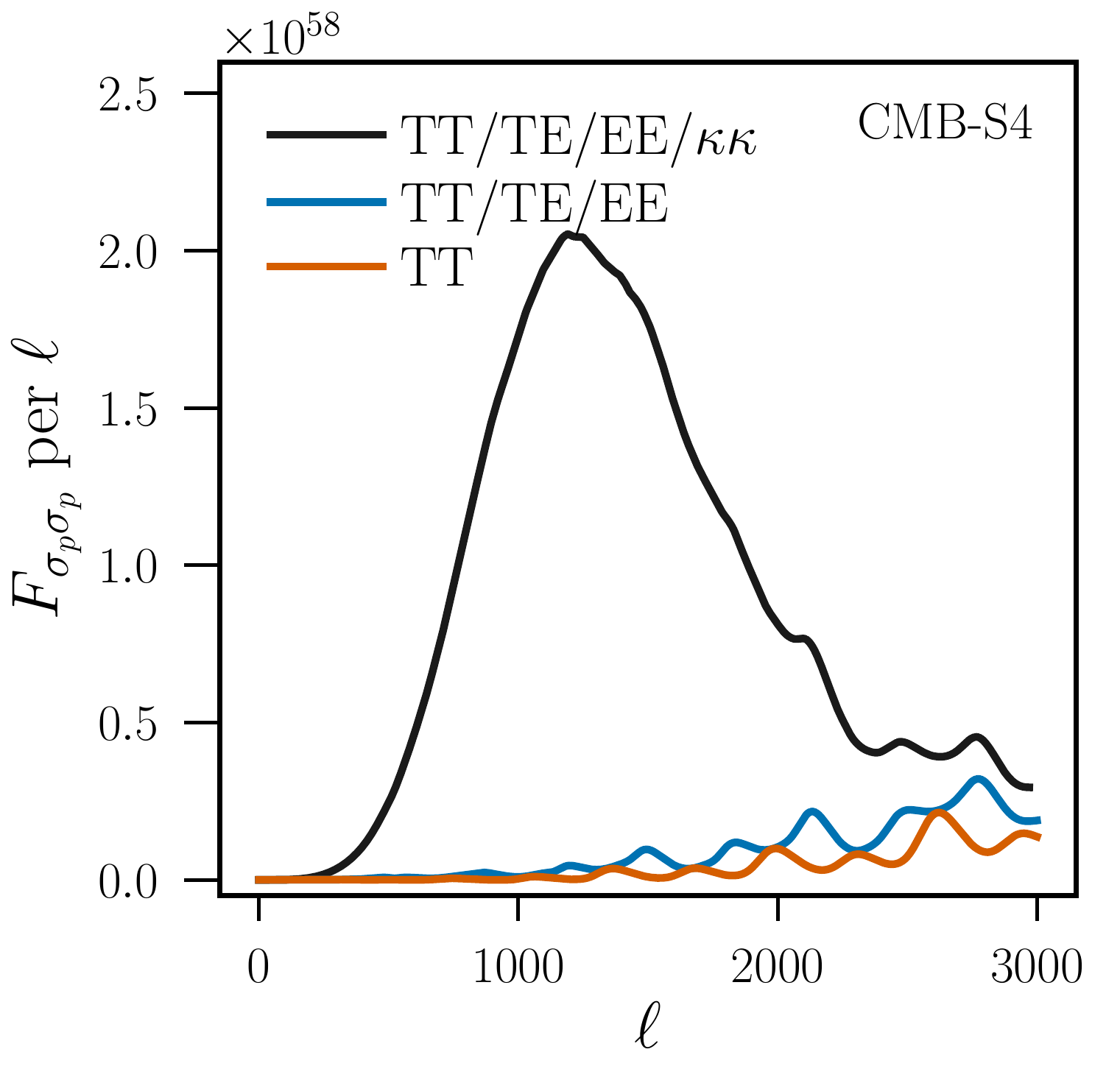}}
\caption{\textbf{[Top:]} Comparison of temperature and polarization contributions toward Fisher information for DM scattering [summands in Eq.~(\ref{eq:fisher})] are shown for an experiment with \textit{Planck} specifications. 
At small angular scales, the suppression of power in temperature dominates the overall DM scattering signal; at large angular scales, the signal comes primarily from shifts in the acoustic peaks and polarization significantly contributes Fisher information. 
\textbf{[Bottom:]} Same as top panel, except for a CMB-S4 experiment. 
In this case, the lensing-convergence power spectrum dominates sensitivity at all angular scales. 
\label{fish_fig_compare_planck}}
\end{figure}

To further quantify the information captured by lensing of the primary CMB, we repeat our Fisher analyses without $\kappa\kappa$, first using only the unlensed and then using only the lensed temperature and polarization power spectra.
We find that there is considerable information in just the lensed CMB, even without the convergence power spectrum.
In particular, for \textit{Planck}, the analysis of lensed power spectra yields a constraint tighter by $\sim$$15$\%, compared to the analysis that uses unlensed spectra.
For CMB-S3, we find a factor of $\sim$$2.6$ improvement, and for CMB-S4, a factor of $\sim$$3.8$.
For \textit{Planck}, forecasts obtained using the lensed CMB primary are within 7\% of those obtained using unlensed CMB combined with the lensing convergence. 
With CMB-S4, the difference reaches a factor of $\sim$2.
From these comparisons, it may seem sufficient to conduct parameter estimation with the lensed CMB alone, in order to capture most of the information on $\sigma_p$. 
However, in Section \ref{subsec:degeneracy} we show that the $C_{\ell}^{\kappa \kappa}$ spectrum carries essential information that breaks degeneracy between DM-baryon scattering and other signatures of new physics.

\subsection{Dependence on Experimental Design} \label{subsec:experimentaldesign}

\begin{figure}
\centerline{
\plotone{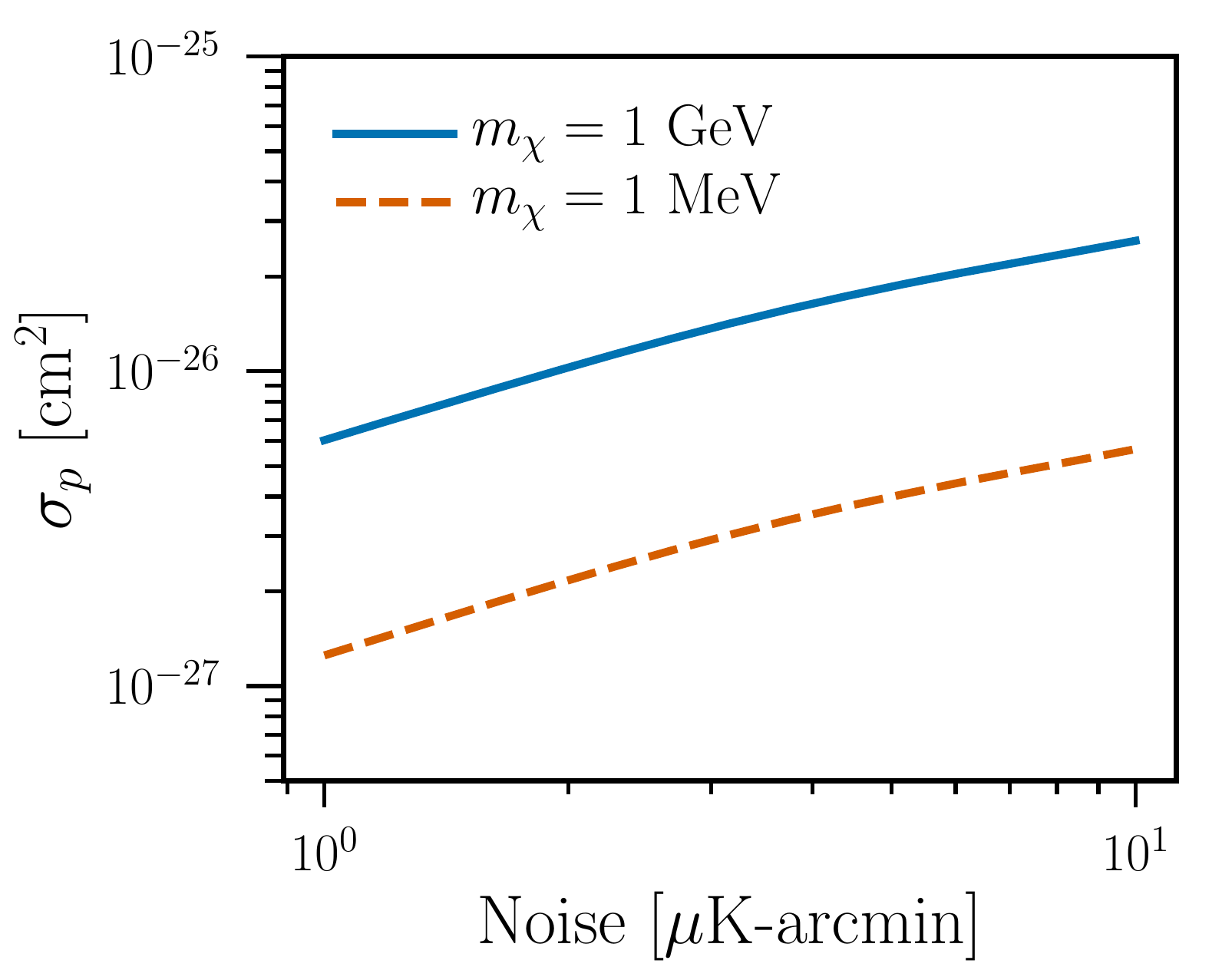}}
\caption{Projected 2$\sigma$ upper limits on the DM-proton scattering cross section as a function of the detector noise level, for a CMB-S4 experiment with a fixed beam width of $\theta_\mathrm{FWHM}$=$1.5$ arcmin, and a fixed sky coverage of $f_\mathrm{sky}$=$0.4$.
Projections are shown for two DM particle masses, as denoted in the legend.  \label{sensitivity}}
\end{figure}
\begin{figure}
\centerline{
\plotone{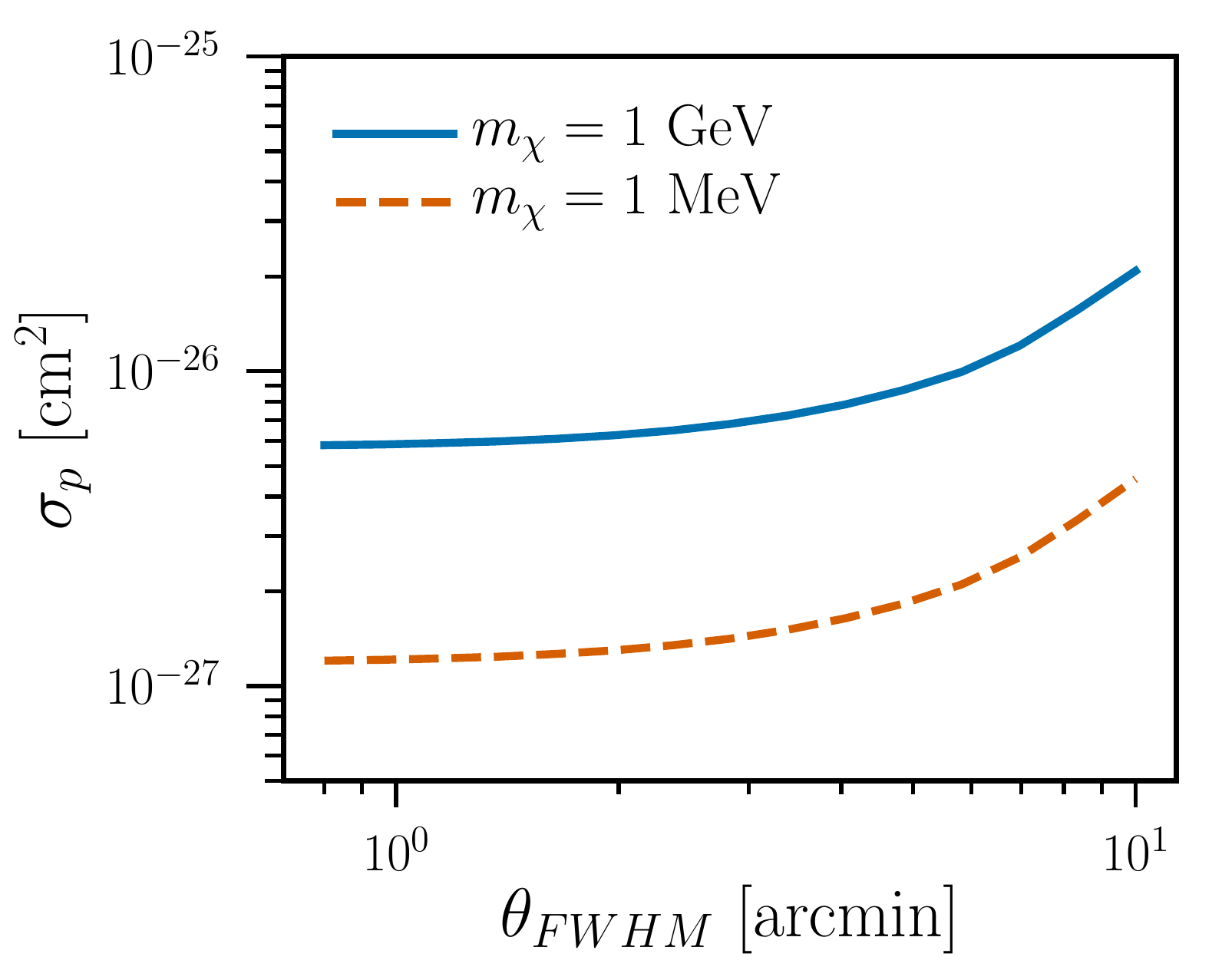}}
\caption{Projected 2$\sigma$ upper limits on the DM-proton scattering cross section as a function of detector beam width $\theta_\text{FWHM}$, for a CMB-S4 experiment with a fixed temperature noise level of $1$ $\mu$K-arcmin, and a fixed sky coverage of $f_\text{sky}$=$0.4$.
Projections are shown for two DM particle masses, as denoted in the legend.   \label{theta}}
\end{figure}
\begin{figure}
\centerline{
\plotone{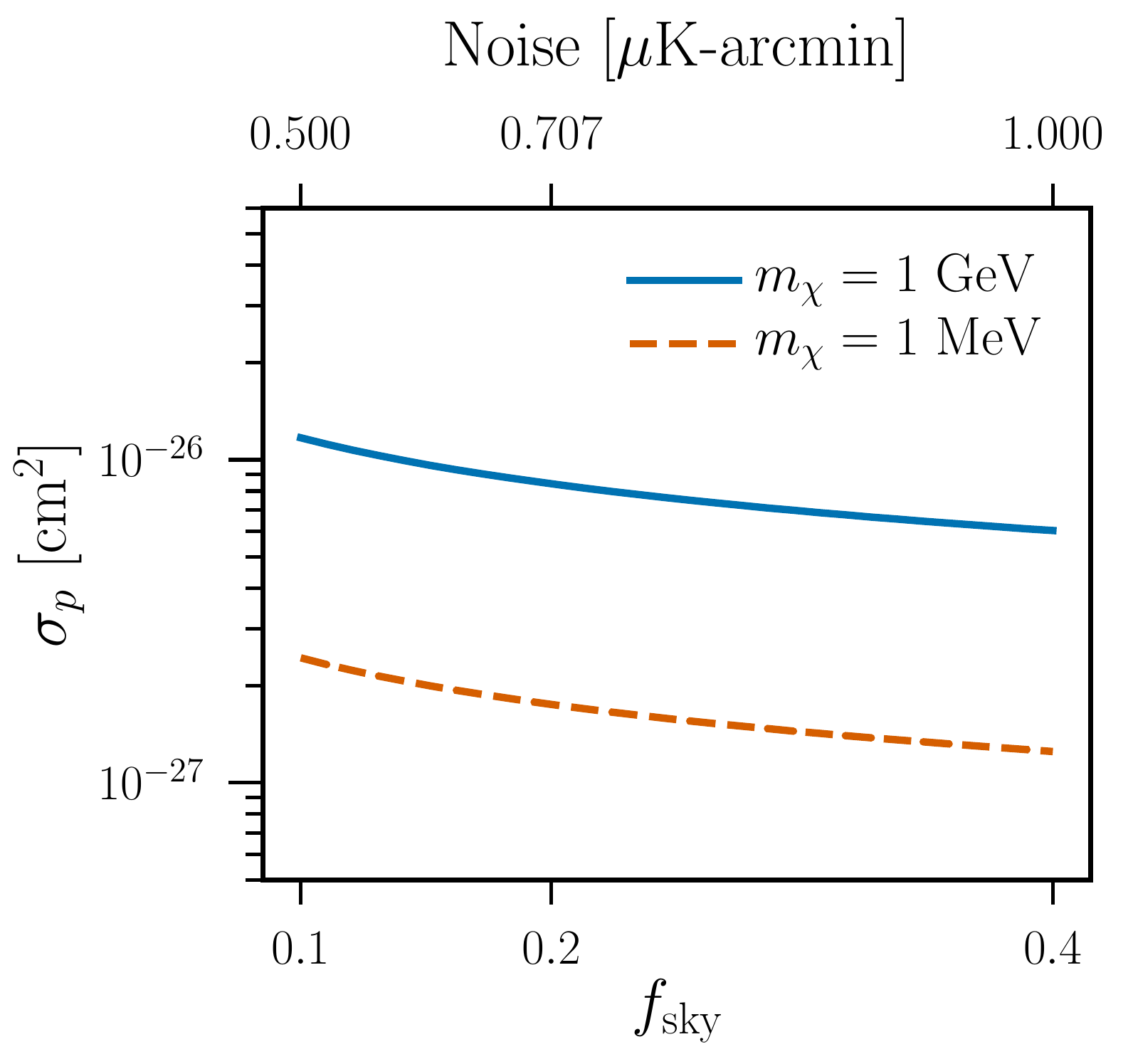}}
\caption{Projected 2$\sigma$ upper limits on the DM-proton scattering cross section for a CMB-S4 experiment, as a function of sky coverage and experimental noise, where the squared detector noise level divided by sky fraction is kept constant, to represent a survey of a fixed duration.
\label{fixed_effort}}
\end{figure}
To understand how sensitivity to DM-baryon scattering depends on the design and survey strategy of future ground-based experiments, we focus on a CMB-S4-like survey.
We first show the projected 2$\sigma$ upper limit on the scattering cross section in Figure \ref{sensitivity}, for a fixed beam size of $\theta_\mathrm{FWHM}=1.5$ arcmin, and fractional sky coverage of $f_\mathrm{sky}$=$0.4$; the projected upper limits are shown for a range of experimental noise levels, assuming white Gaussian noise, for 1 GeV (solid) and 1 MeV (dashed) DM particle mass.
We see that, regardless of the mass, decreasing the noise from 10$\mu$K-arcmin to 1$\mu$K-arcmin secures a factor of $\sim$3 improvement in constraining power (for a fixed resolution and sky coverage).
The dependence of the projected sensitivity on the resolution $\theta_\text{FWHM}$ is shown in Figure \ref{theta}.
For $\theta_\mathrm{FWHM}$ larger than $\sim$6 arcmin, the $TT$ measurement becomes noise dominated at $\ell < 3000$; since the effect of DM-baryon scattering is most prominent at small angular scales, reducing the beam size below this level leads to a large improvement in sensitivity.
For a beam smaller than 5 arcmin, all multipoles considered in our Fisher analyses become signal-dominated, and further reduction in the beam size has no further impact on projected sensitivity; this corresponds to the saturation of the upper limits on the left-hand-side of Figure \ref{theta}.
Finally, in Figure \ref{fixed_effort}, we show 2$\sigma$ limits on the cross-section, for a fixed ratio of squared noise and sky coverage, chosen to represent a fixed survey duration.
The results shown in this Figure demonstrate that wide coverage of the sky benefits searches for DM-baryon interactions more than availability of deep observations of small patches of the sky.

\subsection{Distinguishing signals} \label{subsec:degeneracy}

\begin{figure*}
\centerline{
\epsscale{1.1}
\plotone{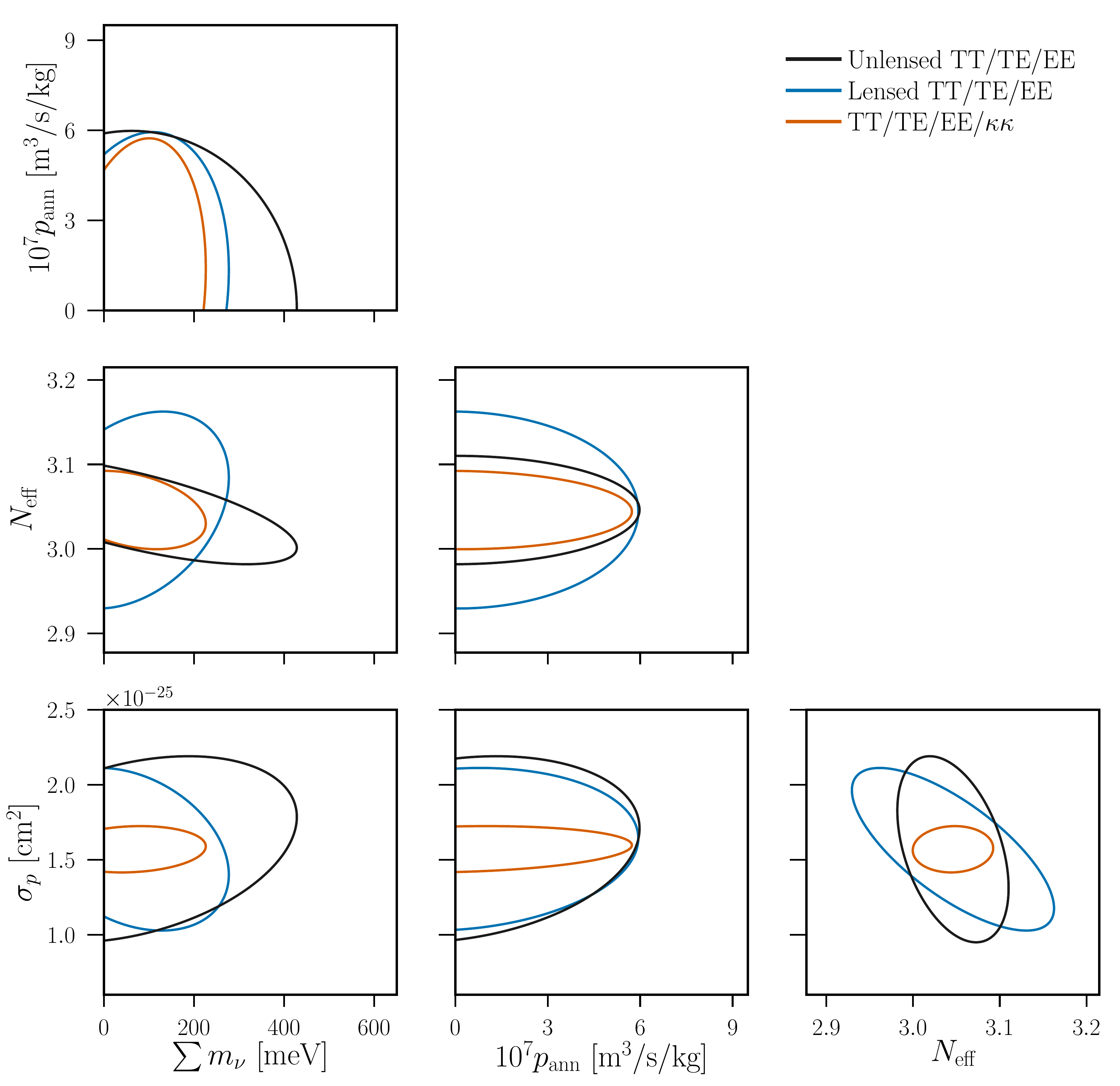}}
\caption{Projected 1$\sigma$ error contours for CMB-S4 experiment, for DM-baryon scattering cross section $\sigma_p$, DM annihilation parameter $p_\text{ann}$ [defined in Eq.~(\ref{eq:ann})], the sum of the neutrino masses $\Sigma m_\nu$, and the effective number of relativistic species $N_{\text{eff}}$.
We assume zero fiducial values for $p_\text{ann}$ and $\Sigma m_\nu$, and set fiducial $\sigma_p$$=$1.6$\times$$10^{-25}$ cm$^2$, close to its current 2$\sigma$ upper limit from \textit{Planck}, fixing the DM particle mass to 1 GeV. 
We set the fiducial value of $N_{\text{eff}}$ to 3.046. 
Different contours correspond to the stand-alone analysis of unlensed (black) or lensed (blue) temperature and polarization power spectra, or to a joint analysis of the unlensed primary anisotropy together with lensing convergence (orange). 
\label{3x_dark_nonzero}}
\end{figure*}

Many signatures of new physics discussed in Section \ref{sec:physics} constitute prime science targets of the next-generation CMB experiments like the Simons Observatory and CMB-S4.
We now examine possible degeneracies between such signals by performing a joint Fisher forecast that includes the standard cosmological parameters (as we did in Section \ref{sec:results}), and the parameters describing DM-baryon interactions $\sigma_p$, DM annihilations $p_\text{ann}$, the sum of the neutrino masses $\Sigma m_\nu$, and the effective number of relativistic species $N_\text{eff}$---all at once.
We assume CMB-S4 experimental parameters given in Table \ref{tab:experiments} and choose a fiducial cosmology with the best-fit \textit{Planck}-2015 $\Lambda$CDM parameters; we set all other parameters to their null-signal values.
The results are shown in Figure \ref{3x_dark_nonzero}, for analysis that includes only the lensed power spectra, only the unlensed power spectra, and both the unlensed power spectra and the lensing convergence power spectrum.

In this Figure, we find a moderate degeneracy between $\sigma_p$ and $\sum m_{\nu}$, when either the lensed, or the unlensed temperature and polarization power spectra are analyzed.
Interestingly, this degeneracy changes direction between the lensed and unlensed spectra. 
In the unlensed spectra, the signals are positively correlated, because their effects on the acoustic peaks are out of phase (see Figure \ref{fig:cartoon_cltteekk}). 
On the other hand, when the lensed spectra are used in the analysis, the dominant effect that drives the constraints is the power suppression at high multipoles.
Since both DM-baryon scattering and the neutrino mass produce similar power suppression at small angular scales, their effects on the lensed CMB power spectra are degenerate.
However, once the lensing convergence power spectrum is included in the analysis, the degeneracy between $\sigma_p$ and $\Sigma m_\nu$ is entirely broken, for CMB-S4 noise levels, as may be expected from Figure \ref{fig:cartoon_cltteekk}.

Furthermore, we find a significant degeneracy between $\sigma_p$ and $N_{\text{eff}}$ in both the lensed and unlensed temperature and polarization power spectra, which is broken once $C_{\ell}^{\kappa \kappa}$ is included into the analysis.
This was expected given that the lensing residuals shown in Figure \ref{fig:cartoon_cltteekk} display a very different scale dependence of power suppression in the two cases: DM-baryon interactions have a progressively larger impact on smaller scales, unlike the relativistic degrees of freedom which equally impact all small scales.
Similarly, inclusion of the lensing convergence alleviates the degeneracy between $N_{\text{eff}}$ and $\sum m_{\nu}$, although to a lesser degree, since the two signals are more alike [see again Figure \ref{fig:cartoon_cltteekk}].
Finally, the power suppression from DM annihilations has no scale dependence and is thus easy to distinguish from all other physical effects we discuss here, using either the primary anisotropy, or the lensing measurements.

\section{Discussion and Conclusions}\label{sec:conclusion}

We presented forecasts for CMB constraints on the spin-independent velocity-independent elastic scattering of DM with baryons, for DM particle masses down to 15 keV; we considered a variety of planned and future CMB experiments.
The scattering signal predominantly appears as a progressively stronger suppression of power on small angular scales in the primary CMB and the CMB lensing power spectrum. 
We found that high-resolution ground-based CMB experiments can deliver a substantial improvement in sensitivity to detecting DM-baryon scattering: CMB-Stage 3 could reduce the current upper limits on the interaction cross section by a factor of $\sim$6 as compared to the results from \textit{Planck}; CMB-Stage 4 could deliver a factor of $\sim$26 improvement; and a CV-limited experiment, a factor of $\sim$200.
While the temperature anisotropy drives the current constraints, future limits will be dominated by the information from the lensing convergence power spectrum measurements (which in turn will be dominated by measurements of CMB polarization). 
In addition, we found that CMB searches for DM-baryon interactions benefit from wide surveys that maximize sky coverage at a fixed survey duration.

Various measurements relating to the neutrinos and the dark sector are core scientific goals of the next-stage CMB experiments. 
Given the challenges for determining the absolute mass scale of neutrinos in laboratory experiments, the first measurement of the sum of their masses, for example, is likely to come from these observations \citep{drexlin2013}. 
Furthermore, DM-baryon interaction strengths and particle masses probed by cosmological measurements lie in a complementary portion of the parameter space outside the lamppost of the most sensitive DM direct detection searches.
It is thus plausible that some of the first signs of new physics beyond the Standard Model may come from cosmological observations.
In light of this, it is important to understand whether different new-physics effects commonly considered in the literature are distinguishable from each other \cite{calabrese2011}.
Exploring this question, we found a moderate degeneracy between DM-baryon scattering, the effective number of relativistic species, and the sum of the neutrino masses, when either only the lensed or only the unlensed $TT$, $TE$, and $EE$ power spectra are analyzed with a Fisher-forecasting method.
This degeneracy is almost entirely broken once the lensing convergence spectrum is considered.
This finding may serve as a guideline for future data analyses: for example, if the DM-proton scattering cross section is non-vanishing, it can bias neutrino mass measurements towards higher values, assuming only the primary anisotropy is considered; put another way, potential inconsistencies in parameter estimation from the primary and secondary anisotropy in future data sets could be a signpost for a complex set of new-physics signals in the CMB data.

Tracers of the large scale structure (LSS) and the matter power spectrum, such as the Lyman-$\alpha$ forest, Baryon Acoustic Oscillations, galaxy cluster counts, and other measurements targeted by the present and future surveys ---DES \citep{DES2005}, LSST \citep{LSST2009}, DESI \citep{DESI2016}---are not considered in this study, but could further advance the sensitivity of cosmological searches for new physics by measuring small scales inaccessible to the CMB.
When considering the LSS observables, it will be particularly important to account for non-linear structure evolution, which we neglected in this work.
While this simplifying choice may have minor impact on forecasts derived using the CMB lensing-convergence power spectra, a careful inclusion of non-linearities using schemes optimized for these modified cosmologies will be essential when considering the LSS observables.
Forecasts for the LSS surveys paralleling those presented in this study for the next-stage CMB, and a consideration of the combined CMB and LSS analyses that will help further break degeneracies between different signals of new physics are a worthwhile exercise left for future work.

\vspace{5mm}

\acknowledgements{VG gratefully acknowledges the support of the Eric Schmidt fellowship 
at the Institute for Advanced Study. 
VG and KB acknowledge KITP and \emph{The Small-Scale Structure of 
Cold(?) Dark Matter} workshop for their hospitality and support under 
NSF grant \#PHY-1748958, during the completion of this work. Computing resources were provided by the PICSciE-OIT TIGRESS High Performance Computing Center at Princeton University. 
The authors are grateful to Nick Battaglia, Jo Dunkley, Joel Meyers, Neelima Sehgal, Blake Sherwin, and Alexander van Engelen for useful discussions.}

\software{CLASS \citep{class},
          Monte Python \citep{audren2013},
          scipy \citep{jones2001},
          astropy \citep{astropy},
          corner.py \citep{foreman-mackey2016}
          }
\appendix

\section{Modified Fisher Forecasting: Priors} 
\label{sec:appendix}

\twocolumngrid

The quadratic dependence of the power-spectra signal on the coupling of the DM particles with protons---a parameter of interest for the main forecasting exercises presented in this work---forced us to modify our Fisher analysis in order to avoid linear dependence of the forecasts on the fiducial value of the coupling and to capture the non-Gaussian shape of the posterior at hand.
We now discuss this new method, which is computationally efficient, while producing posteriors in excellent agreement with outputs from MCMC.
We validated it against the mock-likelihood analyses using the \texttt{Monte Python} code \citep{audren2013}.

The derivative of $C_\ell$ with respect to $c_p$ approaches zero when $c_p$ approaches zero.
This makes it impossible to forecast constraints on the coupling constant using a zero fiducial value, as a zero derivative corresponds to an infinitely poor constraint on $c_p$.
Furthermore, previous data analyses indicated that the posterior for $c_p$ is not Gaussian, which violates the assumptions of the usual Fisher matrix analysis.
The brute-force solution is to make forecasts with Monte Carlo using mock likelihoods, but such methods can be computationally expensive.

We address this problem by introducing a reparametrization that ensures a Gaussian posterior and linear dependence of observables on the parameter---we convert the coupling constant into a cross section $\sigma_p$$\propto$$c_p^2$ (see Section \ref{subsec:DMeff}). 
We then compute the Fisher matrix for the cross section, and then transform back to the coupling by numerically sampling with the covariance matrix and transforming each parameter value appropriately.
Priors can change under this transformation, so the original priors are restored by reweighting the samples. 
The algorithm for this procedure is as follows:
\begin{enumerate}
\item Compute $C_\ell$ derivatives and a Fisher matrix for parameters with approximately Gaussian posteriors; in our case, these are the usual $\Lambda$CDM parameters and $\sigma_p$.
\item Obtain a covariance matrix by inverting the Fisher matrix.
\item Numerically sample a multivariate Gaussian from this covariance matrix.
\item Apply priors by weighting the numerically generated sample of parameters.
\item Transform the sample values of $\sigma_p$ back to $c_p$ in order to estimate marginalized probability density for $c_p$.
\end{enumerate}
Priors are in general not invariant under this reparametrization. 
In our case, a flat prior in $c_p$ translates to a $1/\sigma_p^2$ prior on $\sigma_p$. 
This choice of prior can affect the forecast by a factor of order unity. 
We choose a $1/\sigma_p^2$ prior on $\sigma_p$ in order to reproduce previous analysis on Planck data \citep{gluscevic2017, boddy2018} which used a flat prior on $c_p$.

In order to handle a strictly non-negative cross section, we use a parametrization which reduces to our problem after applying a prior that the cross section is non-negative.
This is necessary since a non-negative cross section violates regularity conditions for the Cramer-Rao bound, as part of parameter space in the neighborhood of the fiducial is missing. 
In our specific case, we choose a reparametrization with a ``cross section'' parameter $\bar{\sigma}_p$ that is allowed to be negative, where negative cross sections are the negative of the effect of positive cross section. 
Defining $\Delta C^{XY}_{\ell}(\bar{\sigma}_p)$ to be the derivative of $C^{XY}_{\ell}$ at the fiducial parameter values, we have
\begin{equation}
\Delta C_{\ell}^{XY} ( \bar{\sigma}_p ) = \text{sgn}(\bar{\sigma}_p) \Delta C_{\ell} ( \sigma_p ).
\end{equation}
When we apply priors, we assign zero weight to samples which have negative $\bar{\sigma}_p$ in order to obtain a constraint on $\sigma_p$.

To complete the discussion of our modified method, we now specify our choices of step size in numerically estimating $\partial C_{\ell}^{XY} / \partial \theta_i$ for the Fisher matrix. 
Steps $\Delta \theta_i$ in a parameter $\theta_i$ must be small enough to accurately estimate $\partial C_{\ell}^{XY}/d\theta_i$, but large enough so that the Boltzmann code can differentiate between $\theta_i$ and $\theta_i + \Delta \theta_i$.
We found that a 1\% step size worked for the $\Lambda$CDM parameters; we list our step size choices in Table \ref{tab:fiducial}.
Our modified version of \texttt{CLASS} has sufficient numerical resolution to use step sizes in the cross section $\sigma_p$ for which the derivatives of the power spectrum converge, but this step size changes for each DM particle mass.
To estimate the step size needed for derivatives to converge, we begin with a large step and iteratively reduce the step until the second derivative of $C_{\ell}$ is sufficiently small. 
We computed this on a grid of DM particle masses, and found the required step size can be described as a power law in the particle mass.
We present this power law in Table \ref{tab:fiducial}.
This choice of step size leads to DM-baryon scattering Fisher constraints which are robust to about 20\% under a factor of two change in step size. 
We also validated these choices by running MCMC on specific masses, and found excellent agreement between out Fisher analyses and MCMC forecasts.
\begin{deluxetable}{c|cc}
\tablehead{
\colhead{Parameter} & \colhead{Fiducial Value} & \colhead{Step Size}
}
\startdata
$\Omega_b h^2$ & 0.0222 & 1\% \\
$\Omega_{\mathrm{DM}} h^2$ &  0.120 &  1\% \\
$\tau_{\mathrm{reio}}$ & 0.06 &  1\% \\
$h$ & 0.69 &  1\%  \\
$A_s$ &  $2.2 \times 10^{-9}$ &  1\%  \\
$n_s$ &  0.966 &  1\%  \\ \hline
$\sigma_p$ (cm$^2$) & 0 & $2 \times 10^{-26} (m_{\chi}/\text{GeV})^{0.15}$ \\
$\Sigma m_{\nu}$ (eV) &  0.06 &  0.02 \\
$p_{\text{ann}}$ & 0 &  $10^{-7}$ \\
$N_\mathrm{eff}$ & 3.046 & 0.08 \\
\enddata
\caption{Fiducial values and step sizes used for the Fisher information matrices computed in this work.
The $\Lambda$CDM parameter step sizes are all set to 1\% of the corresponding fiducial values, but the other fiducial values are set to zero, so we fix their step sizes to finite values. 
The derivatives for the usual $\Lambda$CDM parameters ($\Omega_b h^2$, $\Omega_{\mathrm{DM}} h^2$, $h$, $A_s$, $n_s$) are two-sided, and the derivatives for the extension parameters ($\sigma_p$, $\Sigma m_{\nu}$ , $p_{\text{ann}}$, $N_\mathrm{eff}$) are one-sided and right-handed.
The step for $\sigma_p$ is dependent on DM mass $m_\chi$. \label{tab:fiducial}}
\end{deluxetable}


\end{document}